\documentclass[9pt,twocolumn,twoside]{optica}
\setboolean{shortarticle}{false}
\setboolean{minireview}{false}

\title{Ultrabroadband Nonlinear Optics in Nanophotonic Periodically Poled Lithium Niobate Waveguides}

\author[1,*]{Marc Jankowski}
\author[1]{Carsten Langrock}
\author[2]{Boris Desiatov}
\author[3]{Alireza Marandi}
\author[4]{Cheng Wang}
\author[5]{Mian Zhang}
\author[6]{Christopher R. Phillips}
\author[2]{Marko Lončar}
\author[1]{M. M. Fejer}

\affil[1]{Edward L. Ginzton Laboratory, Stanford University, Stanford, CA, 94305, USA}
\affil[2]{John A. Paulson School of Engineering and Applied Sciences, Harvard University, Cambridge, Massachusetts 02138, USA}
\affil[3]{California Institute of Technology, Pasadena, CA 91125, USA}
\affil[4]{Department of Electrical Engineering, State Key Lab of THz and Millimeter Waves, City University of Hong Kong, Kowloon, Hong Kong, China}
\affil[5]{HyperLight Corporation
501 Massachusetts Avenue, Cambridge, MA 02139}
\affil[6]{Department of Physics, Institute of Quantum Electronics, ETH Zurich, Zurich 8093, Switzerland}
\affil[*]{Corresponding author: marcjank@stanford.edu}

\begin{abstract}
Quasi-phasematched interactions in waveguides with quadratic nonlinearities enable highly efficient nonlinear frequency conversion. In this article, we demonstrate the first generation of devices that combine the dispersion-engineering available in nanophotonic waveguides with quasi-phasematched nonlinear interactions available in periodically poled lithium niobate (PPLN). This combination enables quasi-static interactions of femtosecond pulses, reducing the pulse energy requirements by several orders of magnitude, from picojoules to femtojoules. We experimentally demonstrate two effects associated with second harmonic generation. First, we observe efficient quasi-phasematched second harmonic generation with <100 fJ of pulse energy. Second, in the limit of strong phase-mismatch, we observe spectral broadening of both harmonics with as little as 2-pJ of pulse energy. These results lay a foundation for a new class of nonlinear devices, in which co-engineering of dispersion with quasi-phasematching enables efficient nonlinear optics at the femtojoule level.
\end{abstract}

\setboolean{displaycopyright}{true}

\begin{document}

\maketitle

\section{Introduction}
Phasematched interactions in materials with quadratic ($\chi^{(2)}$) nonlinearities are crucial for realizing efficient second-harmonic generation (SHG), sum- and difference-frequency generation, and optical parametric amplification. These dynamical processes are used as building blocks in many modern optical systems, including near- and mid-infrared light generation \cite{Wang2018,Mayer2016}, ultrashort pulse compression\cite{Bache2007}, supercontinuum generation\cite{Phillips2011}, frequency comb stabilization\cite{Mayer2015}, upconversion detection and quantum frequency conversion\cite{Upconversion2011}, all-optical signal processing\cite{Langrock2006}, coherent Ising machines\cite{CIM2016}, and the generation of nonclassical states of light\cite{SPDC2018}. Weakly-guiding diffused waveguides in periodically-poled ferroelectics like lithium niobate\cite{Ti:LN}, lithium tantalate\cite{PPLT}, and potassium titanyl phosphate\cite{PPKTP} are a commonly used platforms for such devices. These waveguides are conventionally formed by a small refractive index modulation ($\Delta n\sim$0.02) due to in-diffused dopants and exhibit low-loss ($\sim$0.1 dB/cm) modes with field-diameters of $\sim$5 $\mu$m, and quasi-phasematched interactions between these modes through periodic poling of the $\chi^{(2)}$ coefficient. To date, these devices have suffered largely from two limitations. The power requirements of such devices are set by the largest achievable normalized efficiencies ($90\%/$W-cm$^2$ for SHG of 1560-nm light\cite{Langrock2006}), and the phase-matching bandwidths (and hence useful lengths for pulsed interactions) have ultimately been limited by the material dispersion that dominates over geometrical dispersion in weakly-guiding waveguides.

Recent efforts have focused on the development of $\chi^{(2)}$ nanophotonics in platforms such as lithium niobate\cite{Zhang2017}, aluminum nitride\cite{AlNUV} and gallium arsenide\cite{GaAs}. These systems allow for densely integrated nonlinear photonic devices, and achieve efficient frequency conversion due to the large field intensities associated with sub-wavelength mode confinement. The current state of the art of $\chi^{(2)}$ nanophotonic devices comprise two approaches: modal phasematching using the geometric dependence of the phase-velocity of TE and TM modes\cite{LNModal,AlN,GaAs}, and quasi-phasematching using waveguides with periodically poled $\chi^{(2)}$ nonlinearities\cite{Wang2018,Bowers2016}. While modal phasematching has achieved the largest normalized efficiencies to date ($13,000 \%/$W-cm$^2$\cite{GaAs}), the waveguide geometry is determined by the conditions in which the phase velocity of the fundamental and second harmonic are matched. These constraints are lifted in quasi-phasematched waveguides, where the waveguide geometry may be chosen to engineer both the group-velocity and the group-velocity dispersion of the interacting waves. The poling period necessary for quasi-phasematched interactions is then determined by the phase-velocity mismatch in the chosen waveguide geometry. While engineering of these dispersion orders is often done in centrosymmetric waveguides, where the relative sign of the group velocity dispersion and $\chi^{(3)}$ nonlinearity can be chosen to achieve soliton formation and spectral broadening\cite{Mayer2015}, to date there has been no demonstration of dispersion-engineered quasi-phasematched $\chi^{(2)}$ interactions.


In this work we use direct-etched nanophotonic PPLN ridge waveguides to provide the first experimental demonstration of ultra-broadband quasi-phasematched $\chi^{(2)}$ interactions in a dispersion-engineered waveguide. This paper will proceed in three parts: i) We briefly summarize the design and fabrication of nanophotonic PPLN waveguides, ii) we experimentally demonstrate second harmonic generation in a dispersion-engineered PPLN waveguide, and iii) we experimentally demonstrate multi-octave supercontinuum generation in a phase-mismatched PPLN waveguide. The devices shown in section (ii), which have been designed for broadband SHG of wavelengths around 2-$\mu$m, exhibit SHG transfer functions with bandwidths of $\sim$250 nm, and achieve a saturated SHG conversion efficiency in excess of 50\% with pulse energies as low as 60 fJ when pumped with 50-fs-long pulses centered around 2 $\mu$m. The bandwidth and energy requirements of this waveguide represents an improvement over conventional waveguides by 10x and 30x, respectively. In section (iii), we choose the poling period of these waveguides for phase-mismatched SHG, which leads to self-phase modulation with an effective nonlinearity more than two orders of magnitude larger than the pure electronic $\chi^{(3)}$ of lithium niobate. When such a waveguide is driven with pulse energies in excess of a pJ it exhibits a cascade of mixing processes, resulting in the generation and spectral broadening of the first five harmonics. The techniques demonstrated here can be generalized to engineer the transfer functions and interaction lengths of any three-wave interaction based on $\chi^{(2)}$ nonlinearities, and will allow for many of the dynamical processes used in conventional PPLN devices to be scaled to substantially lower pulse energies.

\section{Nanophotonic PPLN Waveguides}
We begin by describing the design and fabrication of nanophotonic PPLN waveguides. A cross-section of a typical ridge waveguide is shown in Fig. 1(a), with the simulated TE$_{00}$ modal field amplitude of the fundamental and second harmonic, respectively. We consider a 700-nm x-cut thin film, and examine the roles of etch depth and waveguide width on the performance of the waveguide. For continuous-wave (CW) interactions, the relevant parameters are the poling period needed to achieve phase-matching, and the effective strength of the interaction. The required poling period for second harmonic generation is given by $\Lambda = \lambda/\left(2n_{2\omega}-2n_\omega\right)$, where $\lambda$ is the wavelength of the fundamental, and $n_{\omega,2\omega}$ is the effective index of the fundamental and second harmonic modes, respectively. The poling period is shown as a function of waveguide geometry in Fig. 1(b), and exhibits a linear scaling in width and etch depth, with larger waveguides having larger poling periods. The typical measure of nonlinearity is the normalized efficiency, $\eta_0$, which specifies the efficiency for phasematched, undepleted, CW SHG in a nonlinear waveguide as $P_{2\omega}/P_\omega = \eta_0 P_\omega L^2$. $\eta_0$ is shown in Fig. 1(c), and scales with the inverse of the area of the waveguide modes,
\begin{figure}[tb]
  \centering
  \includegraphics[width=\columnwidth]{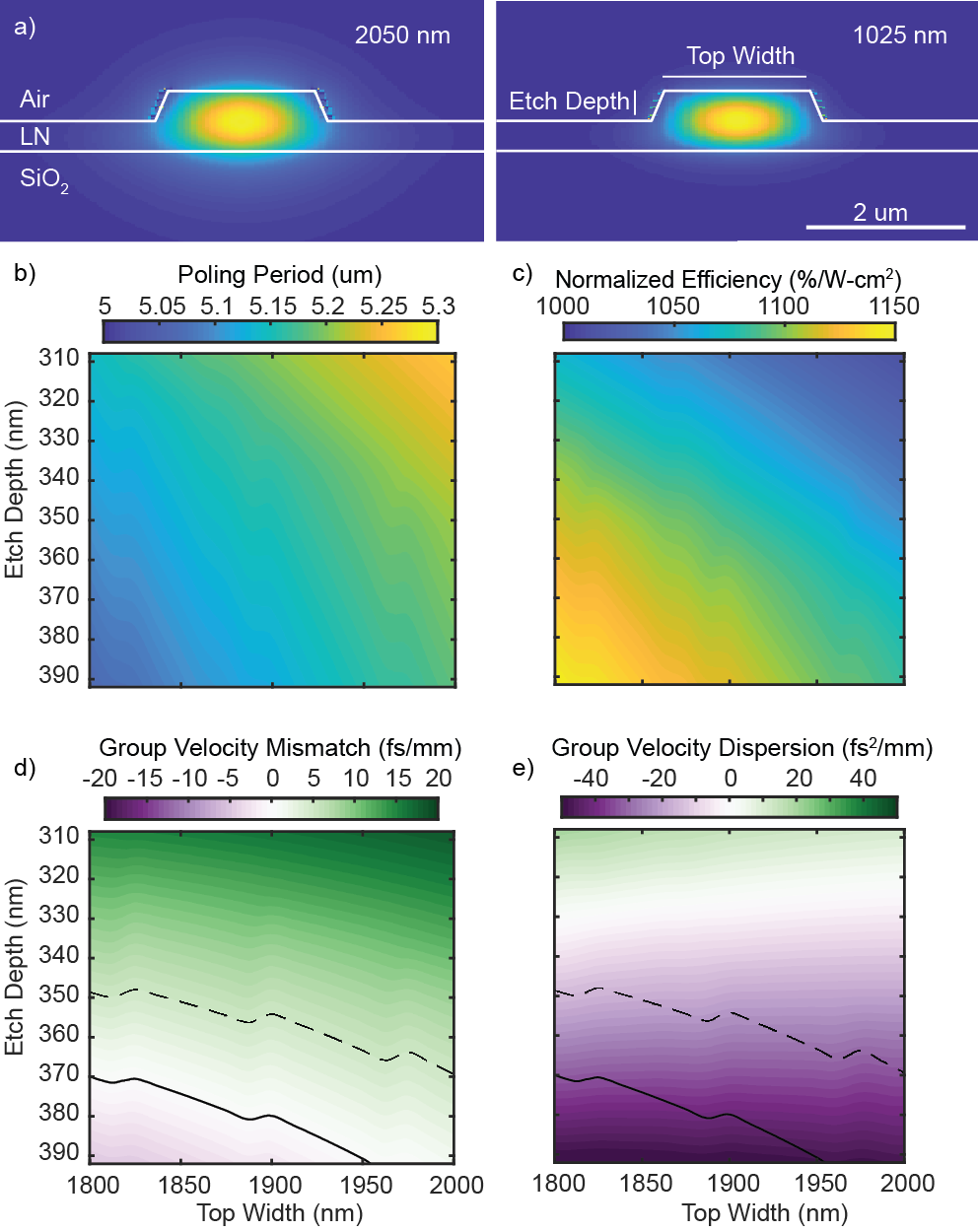}
\caption{a) Waveguide cross-section, showing the normalized electric field associated with the simulated TE$_{00}$ fundamental (left) and second harmonic (right) modes. The waveguides shown here correspond to a top width of 1850 nm, an etch depth of 350 nm, and a starting film thickness of 700 nm. b,c) Simulated poling period and normalized efficiency, respectively, as a function of waveguide geometry. d,e) Simulated $\Delta k'$ and $k_\omega''$, respectively. The solid black line denotes $\Delta k'=0$, and the dashed black contour line shows geometries that achieve $|\Delta k'|<5$fs/mm. The small oscillations observed in these contour lines are a numerical artifact.}
\end{figure}

\begin{equation}
\eta_0=\frac{2\omega^2 d_\mathrm{eff}^2}{n^2_\omega n_{2\omega}\epsilon_0 c^3 A_\mathrm{eff}},
\end{equation}
where $d_\mathrm{eff}=\frac{2}{\pi}d_{33}$ is the effective nonlinear coefficient for quasi-phasematched interactions that have been poled with a 50\% duty cycle, and $d_{33}=20.5$ pm/V for SHG of 2050-nm light. This value is found using a least squares fit to the values reported in \cite{Shoji,Byer} extrapolated to 2 um with constant Miller's delta scaling. $A_\mathrm{eff}$ is the effective area of the interaction and is 1.6 $\mu$m$^2$ for SHG between the modes shown in Fig. 1(a). For a detailed description of  the modal overlap integral involved in computing the effective area, we refer the reader to the supplemental. 

 The role of dispersion will be discussed in more detail in the following sections. We note, for completeness, that the bandwidth of nonlinear interactions is usually dominated by mismatch of the inverse group velocities of the interacting waves, hereafter referred to as the temporal walkoff or group velocity mismatch, $\Delta k'$. In the absence of temporal walkoff, the group velocity dispersion of the fundamental, $k_\omega''$, plays a dominant role. $\Delta k'$ and $k_\omega''$ are shown in Fig. 1(d) and Fig. 1(e) respectively. Temporal walkoff becomes negligible for etch depths $>350$-nm, and anomalous dispersion occurs at wavelengths around 2050-nm for etch depths $>330$-nm.
\begin{figure}[htb]
  \centering
  \includegraphics[width=\columnwidth]{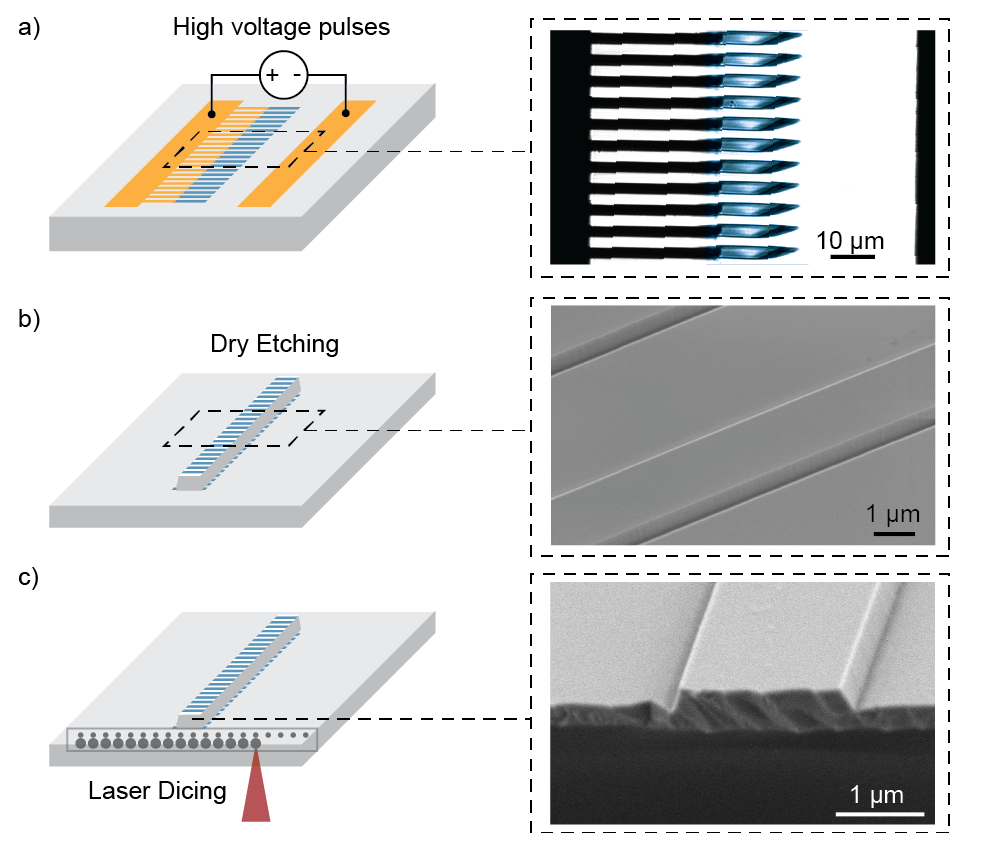}
\caption{a) Schematic of the poling process, resulting in high fidelity domain inversion with a $\sim$50\% duty cycle. b) Waveguides are patterned using an Ar$^+$ assisted dry etch, resulting in smooth sidewalls. c) The samples are prepared using using laser dicing, resulting in optical-quality end-facets.}
\end{figure}

We conclude this section by briefly summarizing the fabrication of the nanophotonic PPLN waveguides used for the remainder of this paper. First, periodic poling is done as described in \cite{Wang2018}. Here, metal electrodes are deposited and patterned on an x-cut magnesium-oxide- (MgO-) doped lithium niobate thin film (NANOLN). Then, several high voltage pulses are applied to the electrodes, resulting in periodic domain inversions (Fig. 2(a)). The inset shows a colorized 2-photon microscope image of the resulting inverted domains with a duty cycle of $\sim$50\%. Second, waveguides are patterned using electron-beam lithography and dry etched using Ar$^+$ ions, as described in \cite{Zhang2017}. This yields low-loss ($<0.1 $-dB/cm) ridge waveguides (Fig. 2(b)). The inset shows a scanning electron microscope (SEM) image of the ridge waveguides, showing smooth sidewalls. Finally, facet preparation is done using a DISCO DFL7340 laser saw (Fig. 2(c)). Here, $\sim$10-$\mu$J pulses are focused into the substrate to create a periodic array of damage spots, which act as nucleation sites for crack propagation. The sample is then cleaved. The inset shows an SEM image of the resulting end-facets, which exhibit $\sim$10-nm facet roughness. The fabricated waveguides have a top width of $\sim$1850 nm, and an etch depth of $\sim$340 nm. The resulting theoretical normalized efficiency is $1100\%$/W-cm$^2$, $\Delta k'=$ 5-fs/mm, and $k_\omega''=$ -15-fs$^2$/mm. The calculated value of $\Delta k'$ is 20 times smaller than that of bulk lithium niobate for 2-$\mu$m doubling, which allows for substantially longer interaction lengths for femtosecond pulses.

\section{Second Harmonic Generation}

In this section we discuss SHG of femtosecond pulses in a nanophotonic PPLN waveguide. We begin by explaining the role of dispersion engineering in phase-matched interactions, and how ultra-broadband phasematched interactions become possible with a suitable choice of waveguide geometry. Then, we describe an experimental demonstration of SHG in a dispersion-engineered PPLN waveguide. The performance of these waveguides, as characterized by the SHG transfer function and normalized efficiency, agrees well with theory and represents an improvement over the performance of conventional PPLN devices, in terms of both bandwidth and normalized efficiency, by more than an order of magnitude.

The coupled wave equations for second harmonic generation of an ultrafast pulse are
\begin{subequations}
\begin{align}
\partial_z A_\omega(z,t) &= -i\kappa A_{2\omega}A_\omega^*\exp(-i\Delta k z)+\hat{D}_{\omega}A_{\omega}\label{CWE1}\\
\partial_z A_{2\omega}(z,t) &= -i\kappa A_\omega^2\exp(i\Delta k z)-\Delta k'\partial_t A_{2\omega}+\hat{D}_{2\omega}A_{2\omega}\label{CWE2}
\end{align}
\end{subequations}
where $A_\omega$ and $A_{2\omega}$ are the complex amplitudes of the modal fields, normalized so that $|A(z,t)|^2$ the instantaneous power at position $z$. $\kappa$ is the nonlinear coupling, $\kappa^2 = \eta_0$, and $\Delta k$ is the phase mismatch between the carrier frequencies, $\Delta k = k_{2\omega}-2k_\omega-2\pi/\Lambda$. The dispersion operator, $\hat{D}_\omega=\sum_{j=2}^{\infty}\left[(-i)^{j+1}k_\omega^{(j)}/j!\right]\partial_t^j$, contains contributions beyond first order, where $k_\omega^{(j)}$ represents the $j$th derivative of propagation constant $k$ at frequency $\omega$.

For SHG in the limit where the fundamental wave is undepleted, these equations may be solved using a transfer function approach\cite{Imeshev_JOSAB2000A,Imeshev_JOSAB2000F}. Here, the response of the second harmonic to the driving nonlinear polarization is computed by filtering the driving polarization with the transfer function for CW SHG. We implement this approach analytically in two steps. First we calculate the second harmonic envelope that would be generated in the absence of dispersion, $A_{2\omega}^\mathrm{ND}(z,t)=-i\kappa A_\omega^2(0,t)z$. Then, the power spectral density associated with this envelope is filtered in the frequency domain, using the CW transfer function for SHG
\begin{equation}
|A_{2\omega}(z,\Omega)|^2=\mathrm{sinc^2}(\Delta k(\Omega) z/2)|A_{2\omega}^\mathrm{ND}(z,\Omega)|^2.\label{TF}    
\end{equation}
Here, $A_{2\omega}(z,\Omega)=\mathcal{F}\lbrace A_{2\omega}(z,t)\rbrace (\Omega)$ is the Fourier transform of $A_{2\omega}(z,t)$, and $\Omega$ is the frequency detuning around $2\omega$. The dispersion of a nonlinear waveguide modifies the bandwidth of the SHG transfer function through the frequency dependence of $\Delta k(\Omega)=k(2\omega+2\Omega)-2k(\omega+\Omega)-2\pi/\Lambda$. In conventional quasi-phasematched devices, the bandwidth of the generated second harmonic is typically dominated by the group-velocity mismatch between the fundamental and second harmonic, $\Delta k(\Omega)\approx 2\Delta k'\Omega$, with corresponding  scaling law for the generated second harmonic bandwidth $\Delta\lambda_\mathrm{SHG} \propto 1/|\Delta k'|L$. As discussed previously, the geometric dispersion that arises due to tight confinement in a nanophotonic waveguide may substantially alter $\Delta k'$. Ultra-broadband interactions become possible when the geometric dispersion of a tightly confining waveguide achieves $\Delta k'=0$. For the waveguides fabricated here, both $\Delta k'$ and $k_\omega''$ are small. In this case the corresponding SHG bandwidth becomes dominated by higher order dispersion, and $\Delta k(\Omega)$ must be calculated using the full dispersion relations of the TE$_{00}$ fundamental and second harmonic modes.

\begin{figure}[htb]
  \centering
  \includegraphics[width=\columnwidth]{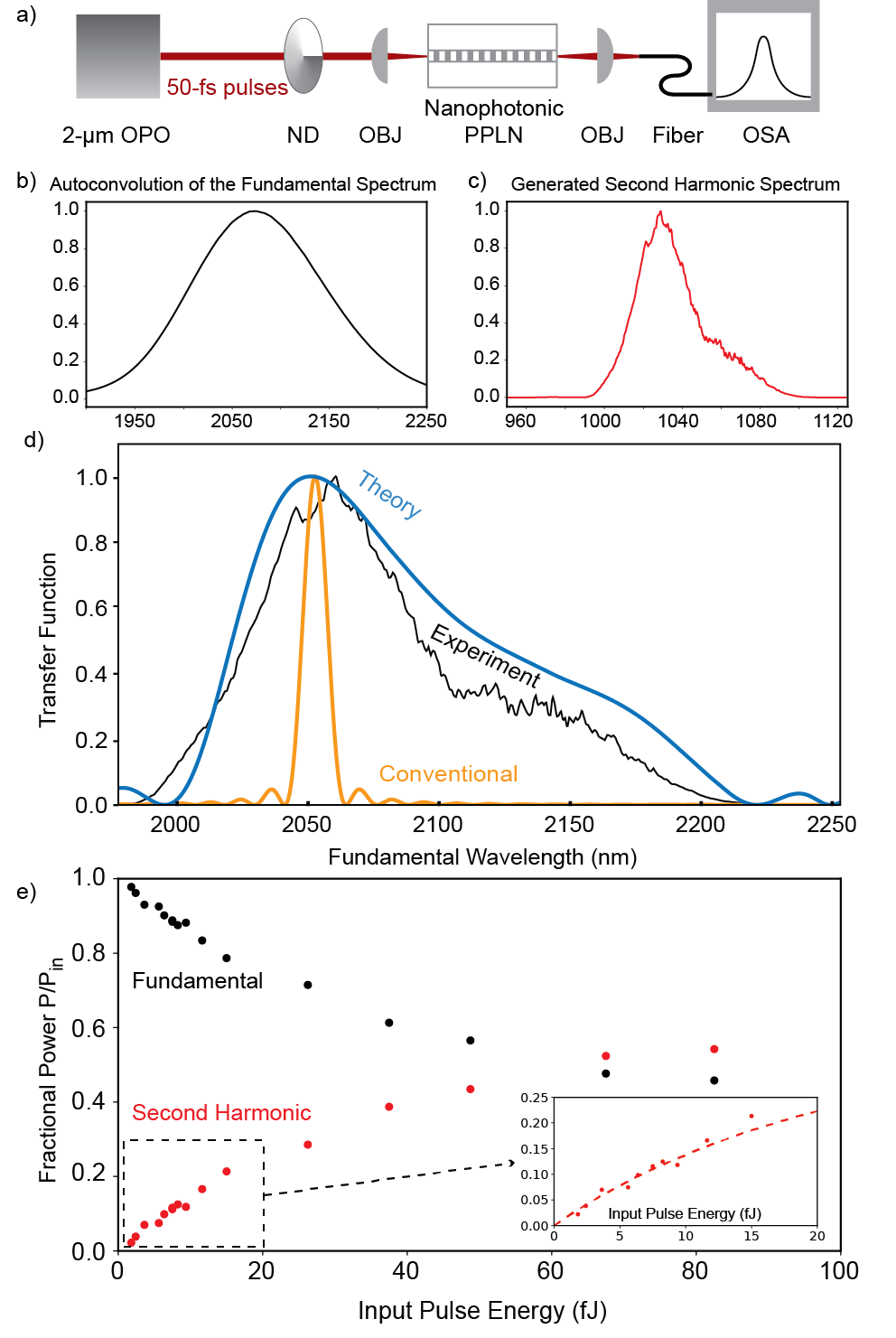}
\caption{a) Schematic of experimental setup. ND- variable neutral density filter, OBJ- reflective objective lens, OSA- optical spectrum analyzer. b,c) Measured spectrum of the driving polarization ($|A_\omega(0,\Omega)*A_\omega(0,\Omega)|^2$) and output second harmonic ($|A_{2\omega}(L,\Omega)|^2$), respectively. d) Measured SHG transfer function (black) for a 6-mm-long nanophotonic waveguide, showing good agreement with theory (blue). The bandwidth of these waveguides exceeds that of bulk PPLN (orange) by more than an order of magnitude. e) SHG conversion efficiency and pump depletion as a function of input pulse energy, showing 50\% conversion efficiency with an input pulse energy of 60-fJ. Inset: Undepleted regime with fit given by Eqn. (\ref{TF}) and a heuristic model for saturation, as described in the text.}
\end{figure}

The experimental setup is shown in Fig. 3(a). We characterize the behavior of 6-mm-long nanophotonic PPLN waveguides using using nearly transform-limited 50-fs-long pulses from a synchronously pumped degenerate optical parametric oscillator (OPO), described in \cite{OPO}. We use reflective inverse-cassegrain lenses (Thorlabs LMM-40X-P01) both to couple into the sample and to collect the output. This ensures that the in-coupled pulses are chirp-free, and that the collected harmonics are free of chromatic aberrations. To characterize the SHG transfer function, we record the spectrum input to the waveguide at the fundamental and output from the waveguide at the second harmonic. Then, we estimate $A_{2\omega}^\mathrm{ND}(z,\Omega)\propto A_\omega(z,\Omega) * A_\omega(z,\Omega)$ using the auto-convolution of the spectrum of the fundamental, shown in Fig. 3(b). The ratio of the measured second harmonic spectrum (Fig. 3(c)) with $A_{2\omega}^\mathrm{ND}$ yields the measured SHG transfer function (Fig. 3(d)), showing good agreement between experiment and theory. These devices exhibit a bandwidth $>$250 nm, when measured between the zeros of the transfer function, which outperforms bulk 2-$\mu$m SHG devices of the same length in PPLN by an order of magnitude. This broad transfer function confirms that the waveguide achieves quasi-static interactions of short pulses across the length of the device. The conversion efficiency of the second harmonic and depletion of the fundamental input to the waveguide is shown as a function of input pulse energy in Fig. 3(e). The inset shows the undepleted regime, denoted by the dotted box in Fig. 3(e). The dotted line is a theoretical fit of Eqn. (\ref{TF}), where we have accounted for a small degree of saturation at the peak of the pulse by using $A_{2\omega}^\mathrm{ND}(z,t)=-i A_\omega(0,t)\tanh{\left(\kappa A_\omega(0,t) z\right)}$. The only fitting parameter used here is a peak CW normalized efficiency of $1000\%$/W-cm$^2$, which agrees well with the theoretically predicted value of $1100\%$/W-cm$^2$, and represents a 45-fold improvement over conventional 2-$\mu$m SHG devices based on proton-exchanged waveguides. When this large CW normalized efficiency is combined with the peak field associated with a 50-fs-long pulse these waveguides achieve 50$\%$ conversion efficiency for an input pulse energy of only 60-fJ, which is a 30-fold reduction compared to the state of the art\cite{fsSHG}.


\section{Supercontinuum Generation}

In this section we discuss spectral broadening by cascaded nonlinearities in a nanophotonic PPLN waveguide. We begin by introducing a heuristic picture based on cascaded nonlinearities in phase-mismatched SHG, and discuss the role of dispersion. Based on this heuristic picture, we show that the effective nonlinearity of these waveguides exceeds than that of conventional $\chi^{(3)}$-based devices, including nanophotonic silicon waveguides. We then describe an experimental demonstration of supercontinuum generation (SCG) in a dispersion-engineered PPLN waveguide. The performance of these waveguides, as characterized by the pulse energies required to generate an octave of bandwidth at multiple harmonics, is an improvement over previous demonstrations in lithium niobate by more than an order of magnitude.

In the limit of large phase-mistmatch, self-phase modulation of the fundamental occurs due to back-action of the second harmonic on fundamental. This can be seen by reducing the coupled wave equations to an effective nonlinear Schrodinger equation for the fundamental wave\cite{Bache2007,Multiscale}. We neglect dispersion beyond second order, and assume the phase-mismatch is sufficiently large to satisfy two criteria: $|\Delta k|\gg\kappa A_0$, where $A_0 = \max(|A_\omega(0,t)|)$, and $|\Delta k|\gg 4\pi|\Delta k'/\tau|$, where $\tau$ is the transform-limited duration of the pulse input to the waveguide. Under these conditions, Eqns.  (\ref{CWE1}-\ref{CWE2}) become
\begin{equation}
\partial_z A_\omega = \frac{ik_\omega''}{2}\partial_t^2 A_\omega+i\gamma_\mathrm{SPM}|A_\omega^2|A_\omega,
\end{equation}
where $\gamma_{SPM}=-\eta_0/\Delta k$. Typically the bounds of $\Delta k$, and thus the strength of the effective self-phase modulation, are set by the temporal walk-off. This constraint is lifted when $\Delta k'\sim$0. For modest values of the phase mismatch ($\Delta k\sim$1 mm$^{-1}$) and the CW normalized efficiency measured previously, the effective nonlinearity is $\gamma_\mathrm{SPM}=100$/W-m. This corresponds to an effective nonlinear refractive index of $n_2=4.8\times 10^{-17}$m$^2$/W. We may compare this to the $n_2$ associated with Kerr nonlinearities in lithium niobate by scaling the values found in \cite{Phillips2011} with a two-band model\cite{TwoBand}. We find $n_2=2.6\times 10^{-19}$m$^2$/W at 2050-nm, which is 185 times weaker than the self-phase modulation due to cascaded nonlinearities. For comparison, typical numbers for $\gamma_\mathrm{SPM}$ based on Kerr nonlinearities in silicon, silicon nitride, and lithium niobate are 38/W-m, 3.25/W-m, and 0.4/W-m respectively\cite{Gaeta2015,Mayer2015,Tang2019}.

In addition to an enhanced nonlinearity, phase-mismatched SHG also generates a spectrally broadened second harmonic. Within the approximations made here the phase-mismatch is constant across the bandwidth of the input pulse, $\Delta k(\Omega)\approx\Delta k(0)$, and the second harmonic is given by
\begin{equation}
A_{2\omega}(z,t) = -i\kappa A_\omega(z,t)^2(\exp(i\Delta k z) - 1)/\Delta k.
\end{equation}
Here, the time varying phase-envelope of the fundamental directly produces a rapidly varying phase of the second harmonic, $\phi_{2\omega}(z,t)\sim2\phi_\omega(z,t)$. Thus, we expect both harmonics to exhibit spectral broadening as the fundamental undergoes self-phase modulation. In practice, the full nonlinear polarization generates a cascade of mixing processes which leads to spectral broadening of several harmonics; a heuristic picture of this process is beyond the scope of this paper.

\begin{figure}[ht]
  \centering
  \includegraphics[width=\columnwidth]{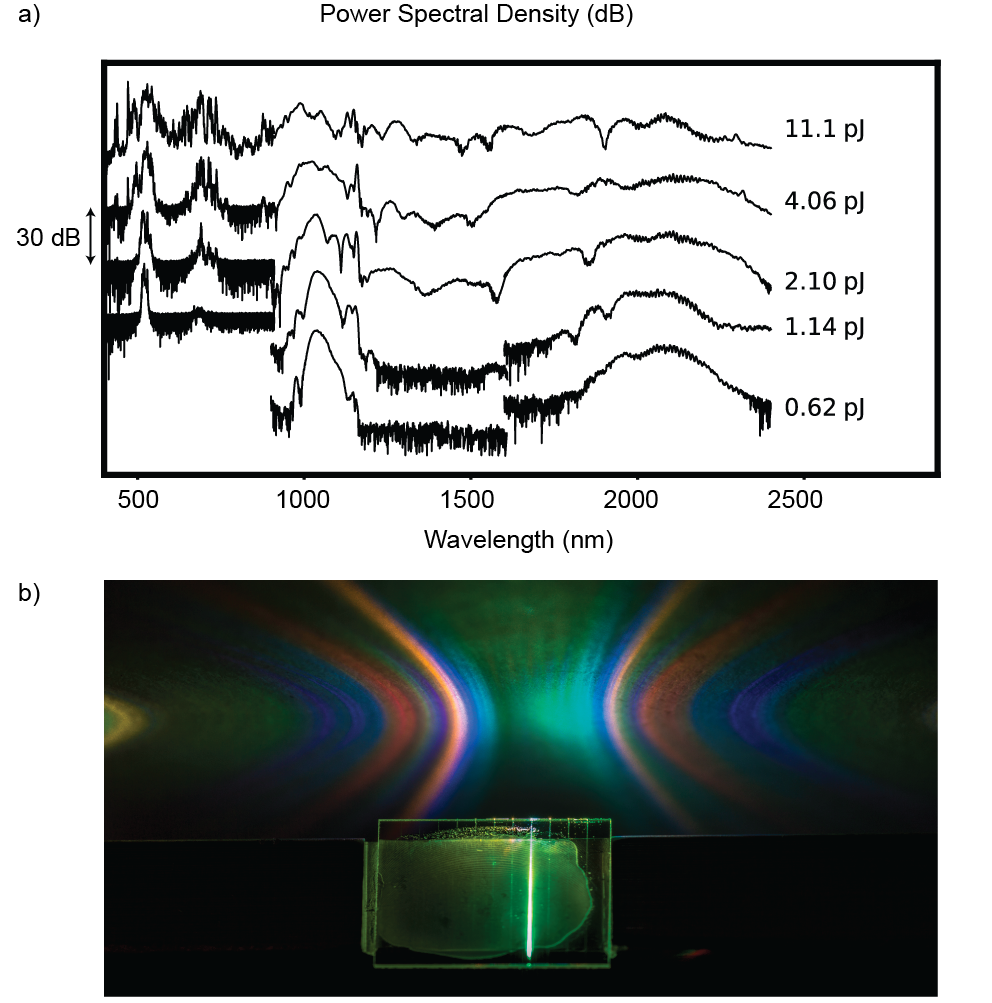}
\caption{a) Evolution of power spectral density over an order of magnitude variation of pulse energy. Adjacent traces are displaced by 30 dB for clarity. The different noise floors correspond to the three spectrometers used. b) Photograph of supercontinuum produced with 11-pJ input to the waveguide.}
\end{figure}

We characterize SCG in a nanophotonic PPLN waveguide with the OPO source and waveguide geometry used in the SHG experiment, however the poling period is now chosen such that $\Delta k L = 3\pi$. We record the output spectrum from the waveguide using three spectrometers: the visible to near-infrared (400-900 nm) range is captured with a Ocean Optics USB4000, the near- to mid-infrared (900-1600 nm) is captured with a Yokogawa AQ6370C, and the mid-infrared (1600-2400 nm) is captured using a Yokogawa AQ6375. The results are shown in Figure 4. The fundamental, second harmonic, and fourth harmonic are observed for input pulse energies as low as 0.5-pJ. For pulse energies >1-pJ, the first two harmonics undergo spectral broadening, and we observe buildup of the third harmonic. As the waveguide is driven with larger pulse energies, all of the observed harmonics undergo spectral broadening. The first two harmonics merge into a supercontinuum spanning more than an octave when driven with 2-pJ of pulse energy. When driven with pulse energies in excess of 10-pJ, the first five harmonics undergo spectral broadening and merge together to form a supercontinuum spanning >2.5 octaves at the -30 dB level. The measured supercontinuum is limited to wavelengths $>400$ nm by the transparency window of our collection optics, and $<2400$ nm by our available spectrometers. A photograph of the multi-octave supercontinuum is shown in Fig. 4(b). The observed diffraction pattern is due to leakage of visible frequencies into slab-modes. The evanescent tails of these modes sample the periodic substrate damage from laser dicing, which acts as a diffraction grating.

To better understand the dynamics and coherence properties of the generated supercontinuum, we simulate Eqns. (\ref{CWE1}-\ref{CWE2}) using the split-step fourier method described in \cite{simulton}, including dispersion to third order. The experimentally measured and simulated spectra output from the waveguide are shown in Figures 5(a) and 5(b), respectively. We note that the simulation includes semi-classical vacuum noise, and the results have been renormalized to have the same peak power spectral density as the experiment in the near-infrared band (900-1600 nm) when driven with a pulse energy of 4-pJ. The two-envelope model used here captures many of the features of the experiment except for the buildup of the higher harmonics, which have been explicitly neglected by only considering $A_\omega$ and $A_{2\omega}$ in the coupled wave equations. The observed spectral broadening agrees well with traditional heuristics derived from the nonlinear Schrodinger equation. If we define the soliton number as $N^2 = \gamma_\mathrm{SPM} U\tau_s /(2k_\omega'')$, where $U$ is the input pulse energy, and $\tau_s = \tau/1.76$, then the the soliton fission length is given by $L_s = \tau_s^2/N k_\omega''$. The soliton fission length approaches the length of the device for an input pulse energy of 1-pJ, which is the energy at which the observed output spectra begin to exhibit spectral broadening. Supercontinuum generation occurs for pulse energies in excess of 2-pJ. Figure 5(c) shows the simulated coherence function, $|g^{(1)}(\lambda,0)|$\cite{Dudley}, which has been calculated using an ensemble average of 100 simulations, for an input pulse energy of 4-pJ (N = 14). The simulations shown here suggest that the spectra are coherent over the range of pulse energies considered, with a calculated $<|g^{(1)}|>=\int|g^{(1)}(\lambda,0)||A(\lambda)|^2 d\lambda/\int|A(\lambda)|^2 d\lambda$ of 0.9996 and 0.9990 for the fundamental and second harmonic, respectively. We note that the approach used here neglects many possible decoherence mechanisms, such as degenerate parametric fluorescence of the third harmonic. Further theoretical and experimental study of the coherence of the supercontinua observed here, and their potential use for $f_\mathrm{CEO}$ detection, will be the subject of future work.



\begin{figure}[t]
  \centering
  \includegraphics[width=\columnwidth]{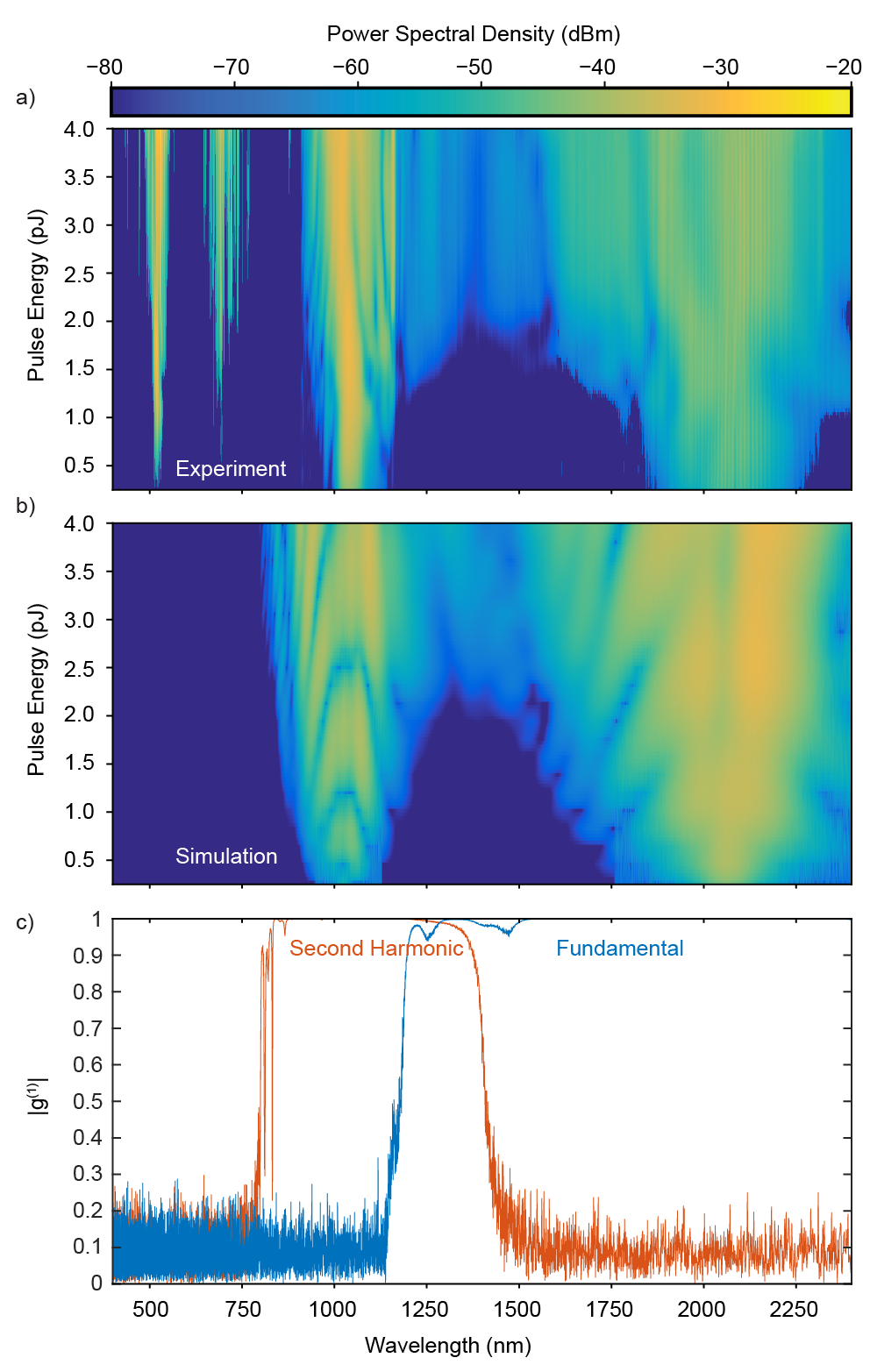}
\caption{Power spectral density output from the chip as a function of input pulse energy a) Experiment, b) Simulation. The power in dBm is measured in 2-nm-wide spectral bins. c) Simulated coherence of the fundamental and second harmonic generated by a 4-pJ pulse, showing $|g^{(1)}|\sim 1$.}
\end{figure}
\section{Conclusion}
We have experimentally demonstrated both SHG and SCG in a dispersion-engineered nanophotonic PPLN waveguide. These waveguides are shown to exceed the performance of current-generation SHG devices by at least an order of magnitude in phasematching bandwidth and pulse energy requirements. Similarly, they achieve self-phase modulation with larger nonlinearities than nanophotonic waveguides based on $\chi^{(3)}$ nonlinearities. These waveguides produce multi-octave supercontinua comprised of multiple spectrally broadenend harmonics with at least an order of magnitude less pulse energy than previous demonstrations in lithium niobate waveguides. These dramatic reductions in energy requirements are made possible by combining the dispersion engineering and large $\eta_0$ available in nanophotonic waveguides with periodically poled $\chi^{(2)}$ nonlinearities. When these techniques are combined they achieve highly efficient quasi-phasematched interactions of femtosecond pulses over long propagation lengths, thereby enabling a new class of nonlinear photonic devices and systems.

\section*{Funding Information}
National Science Foundation (NSF) (ECCS-1609549, ECCS-1609688, EFMA-1741651); AFOSR MURI (FA9550-14-1-0389); Army Research Laboratory (ARL) (W911NF-15-2-0060, W911NF-18-1-0285).

\section*{Acknowledgments}
Patterning and dry etching was performed at the Harvard University Center for Nanoscale Systems (CNS), a member of the NNCI supported by the NSF under award ECCS-1541959. Electrode definition and periodic poling was performed at the Stanford Nano Shared Facilities (SNSF), supported by the National Science Foundation under award ECCS-1542152. The authors thank Jingshi Chi at DISCO HI-TEC America for her expertise with laser dicing lithium niobate.

\section*{Disclosures}
The authors declare no conflicts of interest.




 

\subsection*{See Supplement 1 for supporting content.}

\end{document}


\maketitle

\section{Effective Area of Second Harmonic Generation}
In this section, we derive the nonlinear coupling between two waveguide modes and define the effective area, $A_\mathrm{eff}$, associated with these interactions. The effective area provides a measure of the strength of a nonlinear interaction due to the tight confinement of the waveguide; small effective areas correspond to large field intensities and large normalized efficiencies. We begin by reviewing the necessary components of linear optical waveguide theory to establish the notation used throughout this section. Then, we review the coupled wave equations for second harmonic generation (SHG) in a nonlinear waveguide. The treatment used here accounts for the fully-vectorial nature of the modes, with each field component of the waveguide mode coupled together by the full nonlinear tensor, $d_{ijk}$, of the media that comprise the waveguide. Remarkably, these equations have the same form as the coupled wave equations for SHG in much simpler contexts, such as SHG of plane waves and paraxial gaussian beams. The effective area arises naturally when calculating the normalized conversion efficiency of the power in the second harmonic, $P_{2\omega}/P_\omega$.

Waveguide modes arise as the solution to Maxwell's equations in the absence of a nonlinear polarization, with a dielectric constant that varies in two spatial dimensions,
\begin{subequations}
\begin{align}
\nabla\times H(x,y,z,\omega)&=i\omega\bar{\epsilon}(x,y,\omega)E(x,y,z,\omega),\\
\nabla\times E(x,y,z,\omega)&=-i\omega\mu_0 H(x,y,z,\omega).
\end{align}
\end{subequations}
We note that the media considered here are anisotropic, thus $\bar{\epsilon}(x,y,\omega)$ is a second order tensor. We expand the field in a series of guided modes
\begin{subequations}
\begin{align}
E(x,y,z,\omega) &= \sum_\mu a_{\mu}(\omega)E_{\mu}(x,y,\omega)e^{-ik_{\mu}(\omega) z}\\    
H(x,y,z,\omega) &= \sum_\mu a_{\mu}(\omega)H_{\mu}(x,y,\omega)e^{-ik_{\mu}(\omega) z}
\end{align}
\end{subequations}
where $a_{\mu}$ represents the component of $E$ contained in mode $\mu$ around frequency $\omega$. The transverse mode profiles, $E_\mu$ and $H_\mu$, and their associated propagation constant, $k_\mu$, arise as solutions to an eigenvalue problem\cite{wgmodes}, and each pair of waveguide modes satisfies the orthogonality relation
\begin{equation}
\int_{A}\frac{1}{2}\mathrm{Re}\left(\left[E_\nu\times H_\mu^*\right]\cdot \hat{z}\right)dxdy = P\delta_{\mu,\nu}.\label{orth}
\end{equation}
The fields, as defined, are normalized such that $P=1W$, and therefore the power contained in mode $\mu$ is $P|a_\mu|^2$. For convenience, we define the transverse mode profiles using dimensionless functions $e(x,y)$ and $h(x,y)$
\begin{subequations}
\begin{align}
E_\mu(x,y) = \sqrt{\frac{2Z_0 P}{n_\mu A_{\mathrm{mode},\mu}}}e_\mu(x,y)\label{E_mode},\\
H_\mu(x,y) = \sqrt{\frac{2n_\mu P}{Z_0 A_{\mathrm{mode},\mu}}}h_\mu(x,y)\label{H_mode},
\end{align}
\end{subequations}
where $n_\mu$ is the effective index of mode $\mu$, and $Z_0$ is the impedance of free space. As a consequence of Eqn. (\ref{orth}), the effective area of mode $\mu$ is given by $A_{\mathrm{mode},\mu}=\int(e_\mu\times h_\mu^*)\cdot \hat{z}dx dy$.

The presence of a nonlinear polarization at frequency $\omega$ gives rise to driving terms that cause $a_\mu$ to evolve in $z$. Under these conditions, it can be shown using the methods described in \cite{FejerThesis,UPPE} that $a_\mu$ evolves as
\begin{equation}
\partial_z a_{\mu}(z,\omega) = \frac{-i\omega}{4\mathrm{P}}e^{i k_{\mu}}\int E_{\mu}^* \cdot P_{NL,\mu} dx dy. \label{CWE}
\end{equation}
For second harmonic generation in the limit where one pair of modes is close to phasematching, we consider one mode for the fundamental at frequency $\omega$ and for the second harmonic at frequency $2\omega$ without loss of generality. For the remainder of this article, the modes under consideration will be referred to as $a_{\omega}$ and $a_{2\omega}$ for the fundamental and second harmonic, respectively. In this case, the nonlinear polarization is given by
\begin{subequations}
\begin{align}
P_{NL,\omega} = 2\epsilon_0 d_\mathrm{eff} a_{2\omega}a_{\omega}^*\sum_{jk}\bar{d}_{ijk}E_{j,2\omega}E_{k,\omega}^*e^{-i(k_{2\omega}-k_\omega)z}\label{PNL1}\\
P_{NL,2\omega} = \epsilon_0  d_\mathrm{eff} a_{\omega}^2\sum_{jk}\bar{d}_{ijk}E_{j,\omega}E_{k,\omega}e^{-2ik_\omega z}\label{PNL2}
\end{align}
\end{subequations}
where $i,j,k\in\lbrace x,y,z\rbrace$. $d_\mathrm{eff} = \frac{2}{\pi}d_{33}$ is the effective nonlinear coefficient for a 50\% duty cycle periodically poled waveguide, and $\bar{d}_{ijk}$ is the normalized $\chi^{(2)}$ tensor. For lithium niobate, this is expressed using contracted notation\cite{Nye1985} in the coordinates of the crystal as
\begin{equation*}
\bar{d}_{iJ}=\frac{1}{d_{33}}\left[
\begin{array}{c c c c c c}
0 & 0 & 0 & 0 & d_{15} & d_{16}\\
d_{16} & -d_{16} & 0 & d_{15} & 0 & 0\\
d_{15} & d_{15} & d_{33} & 0 & 0 & 0
\end{array}
\right]
\end{equation*}
where $d_{15}=3.67$ pm/V, $d_{16}=1.78$ pm/V, and $d_{33}=20.5$ pm/V for SHG of 2-$\mu$m light. These values are found using a least squares fit of Miller's delta scaling to the values reported in \cite{Shoji,Byer}, and have relative uncertainties of $\pm 5\%$. We therefore expect a relative uncertainty in any calculated normalized efficiency to be $\pm 10\%$.

We arrive at the coupled wave equations for SHG by substituting Eqns. (\ref{PNL1}-\ref{PNL2}) into Eqn. (\ref{CWE}) and defining $A_\omega=\sqrt{P}a_\omega$
\begin{subequations}
\begin{align}
\partial_z A_\omega = -i\kappa A_{2\omega}A_{\omega}^* e^{-i\Delta k}\\
\partial_z A_{2\omega} = -i\kappa A_\omega^2 e^{i\Delta k}
\end{align}
\end{subequations}
The coupling coefficient, $\kappa$, and the associated effective area are given by 
\begin{subequations}
\begin{align}
&\kappa = \frac{\sqrt{2 Z_0} \omega d_\mathrm{eff}}{c n_\omega \sqrt{A_{\mathrm{eff}}n_{2\omega}}}\\
A_\mathrm{eff}=&\frac{A_{\mathrm{mode,}\omega}^2 A_{\mathrm{mode,}2\omega}}{\left|\int \sum_{i,j,k} \bar{d}_{ijk}e^*_{i,2\omega}e_{j,\omega}e_{k,\omega} dx dy\right|^2}\label{Aeff}
\end{align}
\end{subequations}
We conclude this section by noting that for the waveguide modes considered here the overlap integral in Eqn. (\ref{Aeff}) and the resulting $\kappa$ are real. The general case, in which $\kappa$ is complex, is beyond the scope of this supplemental.